\documentclass[prb,aps,showpacs]{revtex4}
\usepackage{latexsym,epsfig,subfigure,graphicx}

\begin{document}

\title[Time resolved tracking]{Time resolved tracking of a
sound scatterer in a complex flow: non-stationary signal analysis
and applications}
\author{Nicolas Mordant}
\affiliation{\'Ecole Normale Sup\'erieure de Lyon \& {\sc CNRS
umr}~5672, Laboratoire de Physique
\\ 46, all\'ee d'Italie, F-69364 Lyon, France}
\author{Olivier Michel}
\affiliation{Universit\'e de Nice\\ Laboratoire d'astrophysique \&
{\sc CNRS umr}~5525, Parc Valrose, F-06108 Nice, France }
\author{Jean-Fran\c cois Pinton}
\email[e-mail : ]{pinton@ens-lyon.fr} \affiliation{\'Ecole Normale
Sup\'erieure de Lyon \& {\sc CNRS umr}~5672, Laboratoire de
Physique
\\ 46, all\'ee d'Italie, F-69364 Lyon, France}


\date{\today}

\newcommand{\beq}{\begin{equation}}
\newcommand{\eeq}{\end{equation}}
\newcommand{\N}{\vec N}
\newcommand{\X}{\vec X}
\newcommand{\Y}{\vec Y}
\newcommand{\A}{\vec A}
\newcommand{\F}{\vec F}
\newcommand{\vS}{\vec S}
\newcommand{\MS}{{\mbox{\bf S}}}
\newcommand{\MI}{{\mbox{\bf I}}}
\newcommand{\MR}{{\mbox{\bf R}}}
\newcommand{\MP}{{\mbox{\bf P}}}
\newcommand{\MM}{{\mbox{\bf M}}}
\newcommand{\MH}{{\mbox{\bf H}}}
\newcommand{\Tr}{{\mbox Tr}}
\newcommand{\V}{\vec V}

\begin{abstract}
    It is known that ultrasound techniques yield non-intrusive
    measurements of hydrodynamic flows. For example, the study of
    the echoes produced by a large
    number of particles insonified by pulsed wavetrains has led to a
    now
    standard velocimetry technique. In this paper, we propose to extend the
    method to the continuous tracking of one single particle
    embedded in a complex flow. This gives a
    Lagrangian measurement of the fluid motion, which is of importance in
    mixing and turbulence studies. The method relies on the
    ability to resolve in time the Doppler shift of the sound
    scattered  by the continuously insonified particle. \\
    For this
    signal processing problem two classes of approaches are used:
    time-frequency analysis and parametric high resolution
    methods. In  the first class we consider the spectrogram and reassigned
    spectrogram, and we apply it to detect the motion of a small
    bead settling in a fluid at rest. In more non-stationary turbulent
    flows where methods in the second class are more robust, we
    have adapted an Approximated Maximum Likelihood technique coupled
    with a generalized Kalman filter.
\end{abstract}

\maketitle

\section{Introduction}
%---------------------
In several areas of fluid dynamics research, it is desirable to
study the motion individual fluid particles in a flow, i.e. the
Lagrangian dynamics of the flow. The properties of this motion
governs the physics of mixing, the behavior of binary flows and
the Eulerian complexity of chaotic and turbulent flows. Lagrangian
studies are possible in numerical experiments where
chaotic~\cite{Fountain00} and
turbulent~\cite{Yeung89,Squires91,Yeung94,Yeung97} flows have been
studied. For turbulence, the numerical studies are limited to
small Reynolds number flows whose evolution is only followed
during a few large-eddy turnover times.  In addition only the
small scales properties of homogeneous turbulence are captured;
the influence of inhomogeneities (such as large scale coherent
structures) are not taken into account. It does not seem possible
at the moment to extend high resolution turbulent DNS computations
to long periods of time or to high Reynolds number flows.
Experimental studies are thus needed. They differ from the
numerical studies because one cannot tag and follow individual
{\it fluid} particles; most techniques aim at recording the motion
of {\it solid} particles carried by the flow motion. The degree of
fidelity with which solid particles can act as Lagrangian tracers
is an open problem; it depends on the size and density of the
particle. While the interaction between the particle and its wake
can be important for large particles or particles with a large
density difference with the surrounding
fluid~\cite{Maxey83,Sridhar95,Mordant00}, it is generally admitted
that density matched particles with a size smaller that the
Kolmogorov length follow the fluid. Measurements of small particle
motion have been made, using optical techniques that follow
individual particle motion over short
times/distances~\cite{Virant97,Voth98}. We propose here an
acoustic technique that can resolve an individual particle motion
over long periods of time (compared with the characteristic time
of flow forcing).

The principle of the technique is to monitor the Doppler shift of
the sound scattered by a particle which is {\it continuously}
insonified. This is an extension of the pulsed Doppler method that
has been developed to measure velocity profiles and that has many
applications in fluid mechanics and medicine~\cite{Omer99}. The
main advantage of the continuous insonification is to improve the
time resolution of the measurement, although it is limited to the
tracking of a very small number of particles (the tests reported
here are made with only one particle in the flow). The measurement
relies on the ability to track a Doppler frequency and its
variation in time. For this signal processing problem two classes
of approaches have been developed: (i) time-frequency analysis and
(ii) high resolution parametric spectral analysis. Time frequency
methods mainly rely upon the quadratic Wigner-Ville transform, or
smoothed versions of it.  Numerous studies and papers have
recently been published, in which the theoretical issues are
presented (see e.g. the textbooks by Flandrin \cite{Flandrin98} or
Cohen \cite{Cohen95}).  These non parametric techniques are
convenient and well-suited for weakly non-stationary signals with
a good signal-to-noise ratio (SNR). However, time frequency
representations present numerous drawbacks when it comes to
extract trajectory information. Their quadratic nature give rise
to numerous spurious interference terms that require post
processing. For signals with a faster frequency modulation and a
low SNR, we show here that an optimized parametric approach is a
better choice. Parametric high resolution spectral analysis
methods take advantage of an a priori knowledge of the spectral
content of the recorded signal, namely the emitted signal
frequency plus one or many doppler-shifted echoes. Furthermore, a
time-recursive frame for the estimation of the Doppler shift is
proposed here, where the evolution of the frequency is taken into
account in the algorithm.

The two methods are tested in two experiments, in which the
acoustic signals have different time scales and noise levels. The
first experiment is a study of the transient acceleration of a
heavy sphere settling under gravity in a fluid at rest. In this
case the characteristic time scale of velocity variations is slow
($\tau \sim 50$~ms) and the signal to noise ratio is fair (about
20 dB); we show that a technique of reassignment of the
spectrogram gives good results. The second experiments deals with
the motion of a neutrally buoyant sphere embedded in a turbulent
flow. In this case, velocity variations occur over times of about
1~ms and the signal to noise ratio is low (less than 6~dB). We
show that the AML parametric method yields very good results in
that situation.

The paper is organized as follows: in section~\ref{acsec} we
present the acoustic technique and measurement procedure. In
section~\ref{sigsec} we describe the signal processing techniques,
with a particular emphasis on the AML method which has been
developed and optimized to this particle tracking problem.
Examples of applications to measurements in real flows are given
in section~\ref{expsec}.

\section{Acoustical set-up}
%--------------------------
\label{acsec}

\subsection{Principle of the measurement}
%----------------------------------------
In the experimental technique proposed here, a particle is
continuously insonified. It scatters a sound wave whose frequency
is shifted from the incoming sound frequency due to the Doppler
effect. This Doppler shift is directly related to the particle
velocity ${\mathbf v}_{p}$:
\begin{equation}
    \Delta \omega={\mathbf q}\cdot{\mathbf v}_{p} \; \; ,
\end{equation}
where ${\mathbf q}$ is the scattering wavevector (the difference
between the incident and scattered wavevectors ${\mathbf q} =
{\mathbf k}_{\rm scat} - {\mathbf k}_{\rm inc}$) and $\omega$ is
the wave pulsation.

\begin{figure}
    \includegraphics[height=5cm]{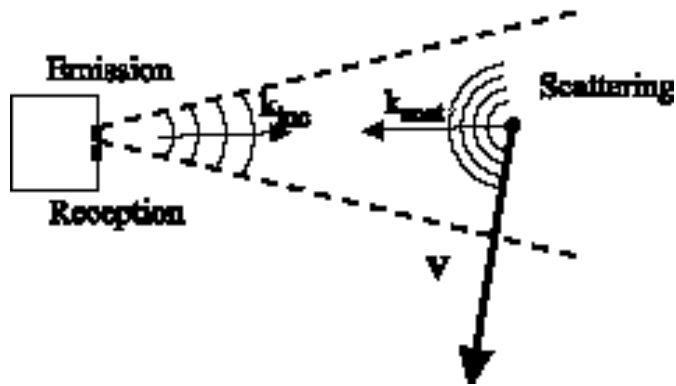}
    \caption{Principle of measurement.
    A large 3D measurement zone is achieved by using a transducer
    size of a few wavelengths. }
    \label{principe}
\end{figure}

We choose a backscattering geometry (see figure~\ref{principe}) so
that ${\mathbf q} = -2 {\mathbf k}_{\rm inc}$ and the frequency
shift becomes
\begin{equation}
    \Delta\omega(t)= - 2\frac{v(t)}{c}\omega_{0} \; \; ,
\end{equation}
where $c$ is the speed of sound, $\omega_{0}$ is the incident
pulsation, and $v(t)$ is the component of the velocity on the
incident direction at time $t$.  We continuously insonify the
moving particle and record the scattered sound. If need be, the
particle position can be obtained by numerical integration of the
velocity signal.

\subsection{Transducers characteristics and acquisition}
%---------------------------------------
We use a Vermon array of ultrasonic transducers made of individual
elements of size 2$\times$2~mm each, separated by 100~$\mu$m.
Their resonant frequency is about 3.2~MHz and their bandwidth at
-3~dB is 1.5~MHz.  Sound emission is set at 3~MHz or 3.5~MHz;
experiments are performed in water so that the wavelength is
$\lambda$=0.50~mm or 0.43~mm.  The corresponding emission cone for
each $d=2$~mm square element is $29^{\circ}$ at 3~MHZ and
$24^{\circ}$ at 3.5~MHz.  In our measurements, the particle to
transducer distance lies between 5~cm and 40~cm, so that
measurements are made in the far field ($d^{2}/\lambda > 10$~mm).
Given maximum flow and particle velocities of the order of
1.5~m.s$^{-1}$, we expect a maximum sound frequency shift of the
order of 5~kHz or 6~kHz, depending on whether the emission is at
3~MHz or 3.5~MHz.  This yields a frequency modulation rate of at
most 0.25\%. One element of the transducer array is used for
continuous sound emission and another for scattered sound
detection. As the operation is continuous (as opposed to pulsed)
and the elements are located close to one-another, we observe a
coupling between the emitter and the receiver of the order of
60~dB (this is due both to electromagnetic and acoustic surface
waves cross-talk).

%\subsection{Acquisition scheme}
%------------------------------

The sound scattered by the moving particle is detected by a
piezoelectric transducer. Upon connection to a 50~$\Omega$
impedance, it yields an electrical signal of about 2 to 30~$\mu$V.
In comparison, the noise is 1~$\mu$V and the electromagnetic
coupling with the emitter is 8~mV. Hence the signal to noise ratio
is between 0~dB and 30~dB.
\begin{figure}
    \subfigure
    {
    \includegraphics[height=5cm]{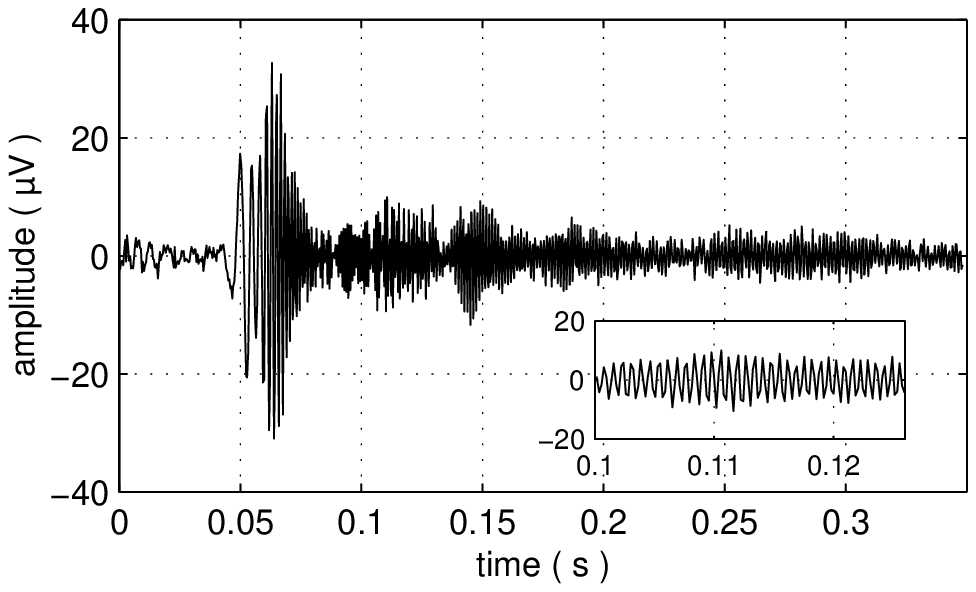}
    }
    \subfigure
    {
    \includegraphics[height=5cm]{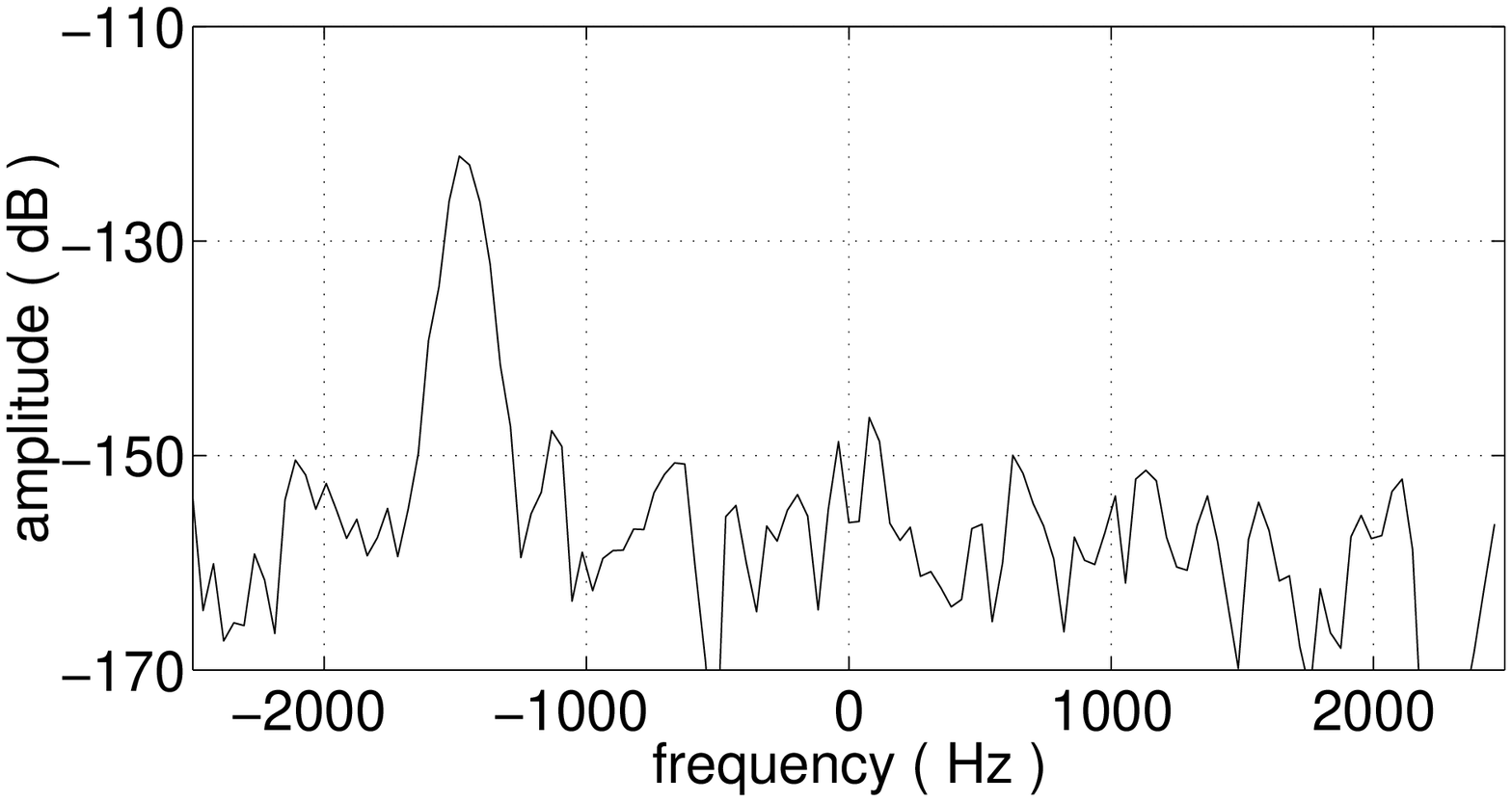}
    }
    \caption{Data from a steel bead (diameter 1~mm) settling in water
    at rest. (a) Typical time series; (b)power spectral density of the inset figure.
    On the x-axis, zero corresponds to the emission frequency.
    }
    \label{sig}
\end{figure}
The transducer output is sampled at 10~MHz over a 21~bit dynamical
range (input range 31.25~mV) and numerically heterodyned at the
emitting frequency.  Then it is decimated at the final sampling
frequency of 19531~Hz.  The acquisition device is a HP-e1430A VXI
digitizer.

\subsection{Scattering by an elastic sphere}
%-------------------------------------------
The study of sound scattered by a fixed solid sphere is a classic
but continuing area of study and difficulties arise in the
interpretation of observed phenomena especially when trying to
deal with elasticity and
absorption~\cite{Gaunard83,Thorne94,Hefner00}. Complex behaviour
is observed linked with resonances of Rayleigh waves at the
surface of the sphere. As a consequence the scattered pressure
distribution varies both in directivity and amplitude.  A generic
expression for the far field pressure is the following :
\begin{equation}
p_{scat}(r,\theta)=p_{inc}\frac{af(ka,\theta)}{2r}e^{ikr} \; \; ,
\end{equation}
where $r$ is the distance from the center of the sphere, $a$ its
radius, $p_{inc}$ the incident pressure on the sphere, $k$ the
incident wavenumber in the fluid, $\theta$ the scattering angle
and $f$ is a form function which depends on the physical
properties of the solid medium. Under very general assumptions,
$f$ can be developed as a series of partial waves:
\begin{equation}
f(ka,\theta) = \frac{2}{ika}\sum_{n=0}^{\infty}
(2n+1)\frac{B_{n}(ka)}{D_{n}(ka)}P_{n}(\cos\theta) \; \; ,
\end{equation}
where $P_{n}$ is the a Legendre polynomial, $B_{n}$ and $D_{n}$
are determinants of matrices composed of spherical Bessel and
Hankel function and their derivatives~\cite{Gaunard83}.
Physically, $f$ represents the sum of the specular echo and of
interferences due to the radiation by Rayleigh
waves~\cite{Thorne94,Hefner00}. As a result, $f$ is a strongly
varying function, particularly for high values of $ka$. In our
experiments we used spheres of different material (polypropylene
PP, steel, tungsten carbide, glass) with corresponding $ka$
between 7 and 15.  The flow acts on the sphere motion, thus
causing its acceleration and, eventually, its rotation. These
effects may change the radiation diagram: first there is Doppler
shift for the sound received by the sphere, and, perhaps more
importantly, the sphere rotation may change the Rayleigh emission.
For these reasons, the evolution of the amplitude of the scattered
sound during the particle motion is quite complex. However, the
observed amplitude modulation (see figures~\ref{sig} and
\ref{sigklac}) varies slowly enough to allow a correct estimate of
the frequency modulation of the scattered sound.

%\clearpage
\section{Signal processing}
%--------------------------
\label{sigsec} Numerous spectral estimation techniques are based
on the ideas behind Fourier analysis of linear time invariant
(LTI) differential equations.  These techniques may be divided
into (i)~non-parametric techniques where the basis functions are
implicitly the harmonically related complex exponentials of
Fourier analysis and (ii)~parametric techniques whose task is the
estimation of the parameters of a (sub)set of complex
exponentials. The spectrogram and the reassigned spectrogram
belong the former category, whereas the maximum likelihood and its
approximate form belong to the latter.

\subsection{Time-Frequency analysis}
%-----------------------------------
The most common time frequency distribution (TFD), the
spectrogram, involves a moving time window.  This window attempts
to capture a portion of the signal which is sufficiently
restricted in time so that stationarity and LTI assumptions are
approximately met. To overcome the inherently poor localization in
the time-frequency plane, a method has been proposed by Gendrin et
al.~\cite{Gendrin79}, and extended more recently by Auger and
Flandrin~\cite{Auger95,Flandrin98}.  The idea is to locally
reassign the energy distribution to the local center of gravity of
the Fourier transform. Despite its ability to exhibit clear and
well localized trajectories in the time-frequency plane, this
technique requires an additional image processing step to extract
the TF trajectory. For rapidly fluctuating frequency modulations
and/or low SNR spurious clusters appear which makes this
extraction difficult. The parametric method presented below is
more robust.

\subsection{AML spectral estimation}
%-----------------------------------
\label{AMLsection} This approach is largely based upon maximum
likelihood spectral estimation (see e.g. Kay\cite{Kay88}).  The
fundamentals are briefly recalled, as they serve as a basis for
the approximate likelihood scheme, originally developed by
Clergeot and Tressens \cite{Clergeot90}.  This work is extended
here within a recursive estimation frame, thus allowing to track
the variations of the Doppler frequency shift induced by fast
velocity changes of a scattering sphere imbedded in a turbulent
flow. Michel and Clergeot have developed a similar approach for
non stationary spectral analysis in an array processing
frame\cite{Michel91a,Michel91b}.

\subsubsection{Introduction}

In this section, we address the problem of estimating the
frequencies $f_1,\ldots,f_M$ of $M$ harmonic signals embedded in
noise, from a small number of samples \beq x(t)=\sum_{m=1}^{M}
a_m(t) \exp( j(2\pi f_mt + \phi_m) + n(t)\; . \eeq As the number
of sampling points that are supposed to be available is low,
classical Fourier based approaches fail to provide good results.
We focus our attention on parametric approach, where an a priori
knowledge about the structure of the signal is taken into account
to improve the analysis.

The following assumptions are made:
\begin{itemize}
\item The time series is regularly sampled with time period $T_s$, so
as to insure $\frac{1}{T_s}> \frac{f_{max}}{2}$ where $f_{max}$
stands for the bandwidth of the anti-aliasing filter used in the
recording process.  For convenience, $T_s$ will be set to $T_s=1$
and the term {\em frequency} will refer to {\em normalized}
frequency (i.e. the actual frequency, divided by
$F_s=\frac{1}{T_s}$.
\item The amplitudes $a_i(t)$ are
deterministic but unknown.
\item The noise is a complex white
gaussian circular and iid (independent increment identically
distributed) with (unknown) variance $\sigma^2$; the distribution
function of a $k$-dimensional vector $\N$ defined by \beq \N(t) =
[n(t), n(t+1), n(t+2), \ldots, n(t+(K-1))]^T \eeq reads \beq
p(\N)=\frac{1}{(\sqrt{2\pi} \sigma)^K} \exp \left(
\frac{|\N|^2}{2\sigma^2} \right) \; .\label{Noise} \eeq
Furthermore, the noise and the signal are independent.
\item The term {\em
observation} refers to a set of $Q$ $K$-dimensional vectors
constructed from the sampled time series, according to \beq
\X(t_j) = [x(t_j), x(t_j+1), x(t_j+2), \ldots, x(t_j+(K-1))]^T
\mbox{ \hspace{5mm} } j=1,\ldots, Q \eeq
\item The frequency $f$ is supposed to remain constant during an
observation.

% Note that if the vectors$\X$ are obtained by time-shift over the
% recorded time series, an observation runs over $K+Q-1$ samples,i.e.
% the actual duration of one observation is $T_{obs}=(Q+K-2)T_s$
\end{itemize}

Under the assumption that the noise process is iid, and that the
observed vectors are corrupted by independent realisations of the
noise process, the likelihood of the observation is given by the
product of the likelihood of each vector.  Let $\mathcal P$ be the
set of searched parameters ($\mathcal P$ contains $\sigma^2$,
$F=\{ f_1,\ldots,f_M \}$ and the $ \A=[a_1\exp(j\phi_1), \ldots ,
a_M\exp(j\phi_M]^T$), the loglikelihood of an observation is
simply given by \beq {\mathcal L}({\mathcal
P})=-KQ\log(2\pi\sigma^2)-\frac{1}{2\sigma^2}\sum_{q=1}^{Q}|\X(q)-\MS
(\F) \A (q)|^2 \label{ML}\; , \eeq where \beq \MS(\F) = [\vS_1,
\ldots, \vS_M] = \left(
\begin{array}{cccc}
    1 & \exp(2 \pi f_1)& \ldots &\exp(2 \pi (K-1)f_1 \\
    \vdots & & & \vdots \\
    1 & \exp(2 \pi f_M & \ldots &\exp(2 \pi (K-1)f_M
\end{array}
\right)^T\; . \eeq According to the maximum likelihood principle,
the set $\mathcal P$ of parameters must be chosen in order to
maximize expression (\ref{ML}).

\subsubsection{Reduced expression}
%---------------------------------
Minimizing (\ref{ML}) jointly for all the parameters is usually
untractable. Most authors propose a separate maximisation for each
of the parameters. For our application, the spectral components
(i.e. $\A$ and $\F$) are the relevant variables. We first maximize
with respect to $\A$ and derive an expression for the optimal
$\F$; $\sigma^{2}$ is estimated independently.

The value of vector $\A$ which minimizes the norm $|\X(q)-\MS (\F)
\A (q)|^2$ is easily obtained~: \beq \A(q) = (\MS^+ \MS)^{-1}
\MS^+(\F) \X(q) \; . \label{A} \eeq Note that the ``signal only"
vector $\Y = \X - \N$ appears to be the orthogonal projection of
$\X$ on the signal subspace spanned by the row vectors of $\MS$:
\beq \Y = \MS(\F).\A(q)=\MS((\MS^+ \MS)^{-1} \MS^+(\F) \X =
\Pi_s(\F)\X  \; , \label{Y} \eeq where $\Pi_s(\F)$ stands for the
parametric projector on the signal subspace\footnote{It is
straightforward to establish that $\Pi_s \Y = \Y$ for any vector
$\Y$ lying in the signal subspace. Here `parametric projector'
must be understood as the projector calculated for the vector of
frequencies $\F$ }.  Let $\Pi_n(\F)=\MI - \Pi_s(\F)$ be the noise
subspace, $\MI$ is the identity matrix.  By substituting (\ref{Y})
and using the definition of $\Pi_n(\F)$ in the expression of the
log-likelihood (\ref{ML}), one gets the following simplified
expression to minimise \beq {L}(\F) =
\frac{1}{\sigma^2}\sum_{q=1}^{Q}|\Pi_n(\F)\X|^2 \; . \eeq Using
the properties of the trace operator (hereafter denoted $\Tr$) and
those of the projection matrix $\Pi_n(\F)$, the maximum likelihood
estimation of $\F$ takes the common form : minimize \beq L(\F)
=\frac{Q}{\sigma^2} \Tr \left[ \Pi_n(\F) \hat\MR_x \right] \; ,
\label{TrPiR} \eeq where
% $$  \hat\MR_x = \frac{1}{Q}\sum_{q=1}^{Q} \X(q) \X^+(q) $$
$\hat\MR_x$ is an estimate of the correlation matrix $\MR_x$ of
the vector process $\X(q)$.  Minimizing $L(\F)$ in (\ref{TrPiR})
leads to the exact value $\F_{ML}$ which has the maximum
likelihood.  It is important here to emphasize the following~: if
the vectors $\X$ are obtained by time-shift over the recorded time
series, an observation runs over $K+Q-1$ samples, i.e. the actual
duration of one observation is $T_{obs}=(Q+K-2)T_s$.  In this
case, the observed vectors may not be considered as being
corrupted by independent realisations of the noise process, as
some 'time integration' is performed in the estimation of $\MR_x$.
The consequences and interest of such smoothing have been studied
by Clergeot and Tressens\cite{Clergeot90}, and
Ouamri\cite{Ouamri86}, in the frame of array processing (in this
context, 'time integration' becomes 'spatial smoothing').  In the
remainder of this paper, the development are based on
equation~(\ref{TrPiR}), no matter how $\MR_x$ is estimated; see
appendix for the practical implementation.

\subsubsection{Approximate Max-likelihood}
%-----------------------------------------
\label{AMLsec} Equation (\ref{TrPiR}) is still too complicated to
be solved analytically in a simple way. A minimization can be
easily performed if $L(\F)$ has a quadratic dependance in
$\MS$\cite{Clergeot90}. Let $\MR_y$ be the correlation matrix of
the signal vectors $\Y(q)$, the assumption that signal and noise
are independent allow to establish the following equalities
\begin{eqnarray}
\hat\MR_x= \MR_y + \hat\sigma^2 \MI \; ,\\ \MR_y= \MS \MP \MS^+ \;
,
\\ \MP = \mathcal E [\A\A^+] \; ,
\end{eqnarray}
where $\mathcal E$ stands for the mathematical expectation.
Substituting in equation~(\ref{TrPiR}) leads to: \beq
L(\F)=\frac{Q}{\hat \sigma^2} \Tr \left[ \Pi_n(\F) \MS \hat \MP
\MS^+ \right] \; . \label{TrPiR2} \eeq Clergeot and
Tressens\cite{Clergeot90} propose a second order approximation of
$L(\F)$: \beq L_{AML}(\F) =\frac{Q}{\hat \sigma^2} \Tr \left[ \hat
\Pi_n \MS(\F) \hat \MP \MS^+(\F) \right] \; , \label{AML} \eeq in
which $\hat \Pi_n$ is estimated by computing the projector spanned
by the $(K-N)$ smallest eigenvalues of the estimated covariance
matrix $\hat \MR_x$.  They prove that this approach leads to more
reliable estimates of $\F$ at low signal to noise ratio (SNR), and
that the minimization of $L_{AML}$ is asymptotically efficient. In
practice, the following set of equations is used
\begin{eqnarray}
\hat \sigma^2 = \frac{1}{K-M}\Tr(\hat \Pi_n \hat \MR_x) \; ,\\
\Pi_s(\F)=\MS(\F) (\MS^+(\F).\MS(\F))^{-1}.\MS^+(\F) \; , \\
\MS(\F).\hat\MP.\MS^+(\F)=\Pi_s(\F)(\hat\MR_x -
\sigma^2\MI).\Pi_s(\F) \; .
\end{eqnarray}

The approximately quadratic dependence of $L_{AML}$ in $\MS(\F)$,
allows a fast convergence of the minimization algorithm by using a
simple Newton-Gauss algorithm: \beq
   \F(k+1) = \F(k) -\MH^{-1}.\vec{\rm grad}(L_{AML})|_{\F = \F(k)}
   \; ,
\eeq where $k$ stands for the iteration step in the minimization
process, $\vec{{\rm grad}}$ and $\MH$ are the gradient and hessian
respectively (see expressions in the appendix).

\subsubsection{Combining new measurements and estimates}
%-------------------------------------------------------
In this section, it is assumed that new measurements do not allow
by itself the derivation of a good estimate.  The variance of such
an estimate varies as $\frac{1}{T_{obs}} \simeq (K+Q-1)^{-1}$,
whereas integrating new measurements to this estimate allow to
derive a better estimation.\\ Let $\hat{\F}(t)$ be an estimate of
$\F$ at time $t$, and ${\mathcal N}(\hat{\F}(t), \Gamma(t))$ its
density, assumed to be normal with variance
$\Gamma(t)$\footnote{This is generally not the case, but this
assumption remains valid as long as the loglikelihood is well
approximated by its second order expansion around $\hat{\F}$.}. If
a linear evolution model is known for $\F(t)$, one has
\begin{eqnarray}
    \F(t+1) = \MM \F(t) + \varepsilon(t) \; , \\
    p_{t+1|t}(\F)=\mathcal N(\MM\hat{\F}(t), \MM\Gamma(t)\MM^+ +
    \MR_{\varepsilon}) \; ,
\end{eqnarray}
where $\MM$ is the evolution matrix; $\varepsilon$ is a
perturbation term, which is statistically independent from $\F$,
and , $\MR_{\varepsilon}$ is its covariance matrix. $p_{t+1|t}$ is
the probability density function that can be derived for time
$t+1$, if the observations are made until time $t$ only.  As such
an evolution equation is usually unknown, $\MM$ will be set to the
identity matrix in the rest of the paper (see
Michel\cite{Michel91b}) for a detailed discussion).  Applying the
Bayes rule over conditional probabilities gives~: \beq
   p_{t+1|t+1}(\F) = \frac{p_{t+1|t}(\F)\, . \, p_{t+1}(\X | \F ) }{p_{t+1}(\X) } \; .
\eeq Noting that $\log (p_{t+1} (\X | \F))$ is the loglikelihood
function for which a reduced expression has been derived in the
previous section, one gets after all reductions and
identifications the simple following expressions
\begin{eqnarray}
\hat{\F}(t+1|t) = \hat{\F}(t) \label{eqa} \; , \\
\Gamma(t+1|t)=\Gamma(t)+\MR_{\varepsilon} \label{eqb}  \; , \\
\Gamma(t+1)^{-1}=\MH + \Gamma(t+1|t)^{-1} \label{eqc}  \; , \\
\hat{\F}(t+1|t+1)=\hat{\F}(t+1)=\hat{\F}(t+1|t)-\Gamma(t+1)^{-1}.{\vec{\rm
grad}}\label{eqd}  \; ,
\end{eqnarray}
where it can be shown that the gradient function has the same
expression as in the previous section.  $\MR_{\varepsilon}$ is an
unknown matrix which will be practically set to $v^2\MI$, where
$v^2$ will be tuned in order to allow the algorithm to take slight
changes in $\F$ into account.  Furthermore, it is interesting that
the set of expression above expresses a generalized Kalman filter
for estimating $\F$ (in the sense that it relies upon second order
expansion of the loglikelihood functions). The statistical
convergence properties and numerical efficiency of these
approaches are described in the work of Michel \&
Clergeot\cite{Michel91a} and Michel\cite{Michel91b}.

\section{Experimental results}
%-----------------------------
\label{expsec} We first describe the simple case of a particle
settling in a fluid at rest. It is well adapted to the reassigned
spectrogram method because the acoustic signal has  a good SNR and
a slow frequency modulation. We show that it allows to extract the
subtle interaction between the falling particle and its wake. We
then study the more complicated case of the motion of a particle
embedded in a turbulent flow, where the dynamics of motion is much
faster and the SNR is poor. We show that the AML method is well
suited.

\subsection{The settling sphere}
%-------------------------------
\subsubsection{Motivation and experimental setup}
%------------------------------------------------
When a particle is released in a fluid at rest, its developing
motion creates a wake.  The particle velocity is then set by the
balance between buoyancy forces and drag, and additional subtle
effects: first, `added mass' corrections because the particles
`pushes' the fluid, and second, a `history' force because the wake
reacts back on the particle.  Formally, one can write the equation
of motion as \cite{Maxey83,Lawrence95,Mordant00}:
\begin{equation}
    \label{EQR}
    (m_{p} + \frac{1}{2}m_{f})
    \frac{{\rm d}{\mathbf v}_{p}}{{\rm d}t} =
    (m_{p}-m_{f}){\mathbf g}
    - \frac{1}{2}\pi
    a^{2} \rho_{f} \left\|{\mathbf v}_{p}\right\|
    {\mathbf v}_{p} c_{D}(Re)
    +{\mathbf F}_{\rm history}\;\;\; .
\end{equation}
where $m_{p}$ is the particle mass, $m_{f}$ is the mass of a fluid
particle of the same size, $v_{p}$ is the particle velocity,
$\mathbf g$ is the acceleration of gravity, $a$ is the sphere
radius, $\rho_{f}$ is the fluid density, $c_D$ is the static
empiric drag coefficient, $Re$ is the Reynolds number
$Re=\frac{2av_p}{\nu}$ and ${\mathbf F}_{\rm history}$ is the
so-called history force.  In this expression, the drag coefficient
is usually obtained from measurement of the forces acting on a
body at rest in an hydrodynamic tunnel.  The history term,
however, is largely unknown.  Analytic expression can only be
derived in the limit of small Reynolds numbers (less than 10) and
cannot be applied for real flow configurations (e.g. multiphase
flows) where $Re \gg 1$.

\begin{figure}
    \includegraphics[height=4cm]{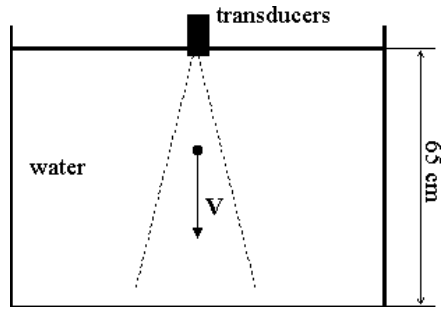}
    \caption{Experimental setup in the
    case of the settling sphere.}
    \label{cuve}
\end{figure}

We perform measurements of the motion of a settling sphere, with
the aim of evaluating the influence of the history forces.  We use
a water tank of size 1.1~m$\times$0.75~m and depth 0.65~m, filled
with water at rest (figure~\ref{cuve}).  The bead is held by a
pair of tweezers, five centimeters below the transducers.  It is
released a time t=0 without initial velocity and its trajectory is
about 50~cm long.  The data acquisition is started before the bead
is released in order to capture the onset of motion.

\subsubsection{Results}
\begin{figure}
    \centering
    \subfigure
    {
    \includegraphics[height=6cm]{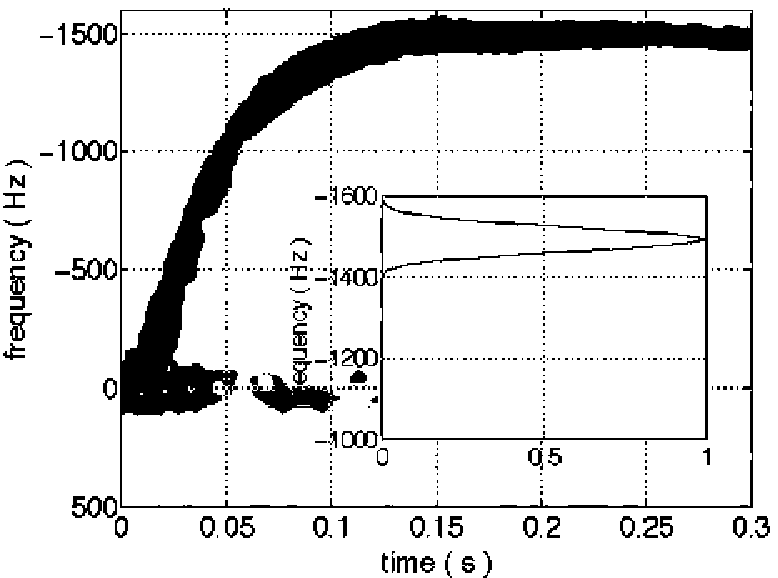}
    }\\
    \subfigure
    {
    \includegraphics[height=6cm]{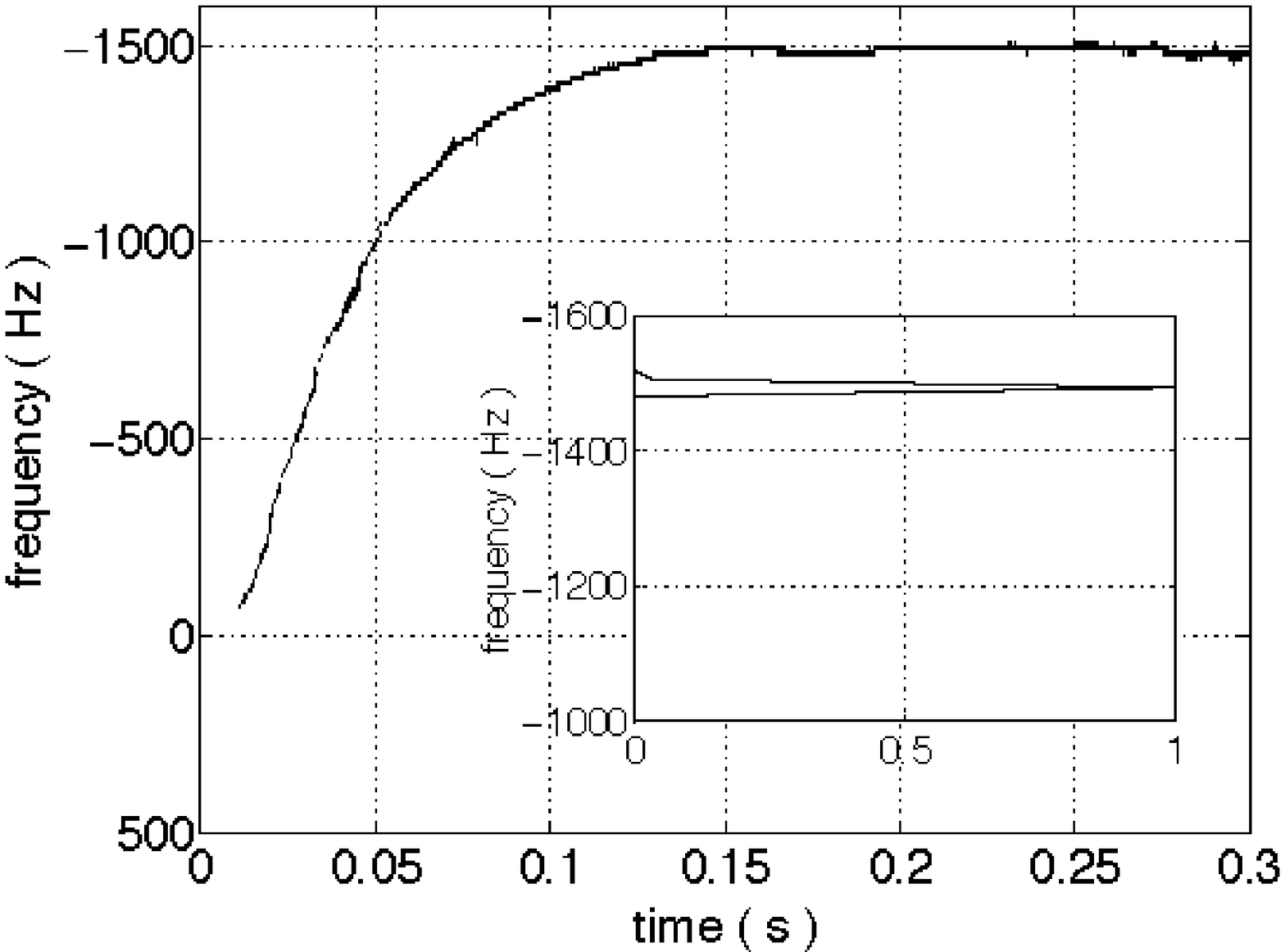}
    }
    \caption{ ({\it a}) Spectrogram of the backscattered sound,
    after heterodyne detection.  ({\it b}) Reassigned spectrogram.  In
    each figure the inset shows a normalized cross-section of the
    spectrogram. The algorithm is that of the {\tt tfrrsp} function of the
    MATLAB time-frequency toolbox~\cite{Matlab}.
    To get rid of the spectral components at zero frequency due
    to the coupling between transducers and at small frequencies
    around zero due to slow motion of the water surface, we use a
    high pass fifth order Butterworth filter of cut-off frequency
    25~Hz (corresponding to a velocity of 5~mm/s). Data of 0.8~mm
    steel bead settling in water at rest. }
    \label{spectro}
\end{figure}

%----------------------
Let us use as a first example, the fall of steel bead, 0.8~mm in
diameter. The Doppler shift during the bead motion is detected
using the spectrogram representation and a subsequent reassignment
scheme. The simple spectrogram and reassigned version are shown in
figure~\ref{spectro}.
The reassignment technique drastically improves the localization
of the energy in the time-frequency plane. In this case, the image
processing step computes $v_{p}(t)$ as the line of maxima. The
precision of the overall measurement depends on two factors~:
first on the intrinsic precision of the reassignment method and
second on the dispersion of the measurements (the reproducibility
of the bead motion over several experiments).  The intrinsic
precision of the reassignment method has been empirically studied
using synthetic signals modelling the particle dynamics plus a
noise that mimics the experimental data. We observed that for our
choice of parameters (a time-frequency picture with 256$\times$256
pixels) the $rms$ precision is about one half pixel both in time
and frequency directions.  The method thus allows a precise
analysis of the dynamics of the fall; we describe below two sets
of experiment that illustrate the potential of the reassignment
technique.

\begin{figure}
    \centering
    \includegraphics[height=6cm]{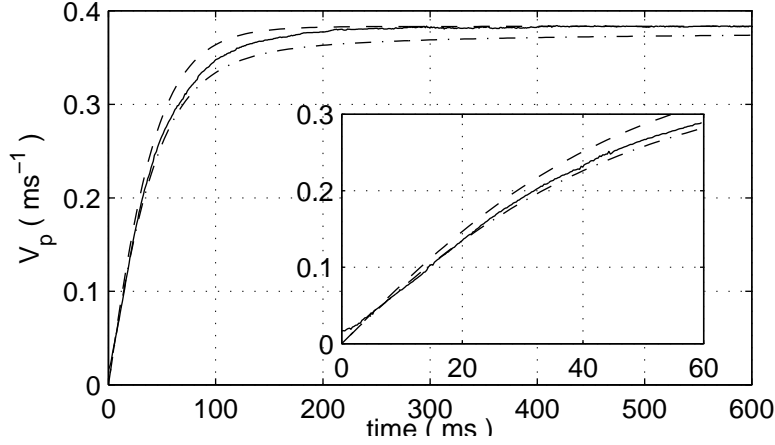}
    \caption{Velocity measurement of a steel bead of diameter 1~mm (solid line),
    compared to numerical simulations without memory force (dashed) and with
    Stokes memory (dash-dotted). The inset
    shows an enlargement near the onset of motion. The Reynolds number, based
    on the limit velocity is 430. The sphere velocity profile  results from
    averaging $n=10$ successive experiments. }
\label{ac1}
\end{figure}

First, we show in figure~\ref{ac1} the velocity of a 1~mm steel
bead (average over ten falls) together with two numerical
simulations based on equation~\ref{EQR}, first without the memory
force and second with the expression of the memory force derived
at low Reynolds numbers (called the Stokes memory term, as in
Maxey \& Riley\cite{Maxey83}). The precision of the detection
technique is sufficient for the measured profile to be compared to
the simulated curves and to draw physical conclusions about the
hydrodynamical forces.  At early times, the trajectory is close to
the simulation with memory force. This is due to the diffusion
away from the bead surface of the vorticity generated at the
boundary~\cite{Maxey83,Lawrence95,Mordant00}. However, as the
instantaneous Reynolds number increases, the curve deviates from
this simple regime: vorticity is advected into the wake.  Memory
is progressively lost and the sphere reaches a terminal velocity
in a finite time as does the simulation without memory.

\begin{figure}
    \includegraphics[height=6cm]{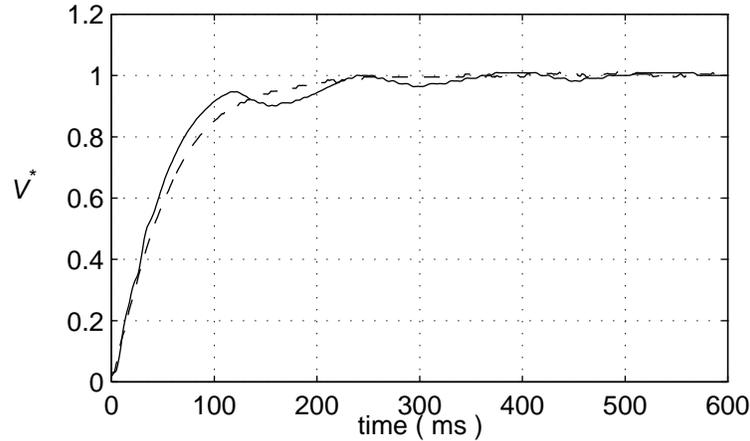}
    \caption{Fall of a tungsten
    carbide sphere D=1~mm (dashed) compared to a glass bead D=2~mm
    (solid), at $Re\sim 400$. The velocity is non-dimensionalized by the limit velocity.
    Curves are not averaged over several experiments.}
    \label{tu1ve2}
\end{figure}

The measurement and signal processing techniques are then tested
on a more non-stationary motion, as in the case of a bead whose
density is close to that of the fluid. In this situation a
stronger interaction is expected between the particle motion and
the development of its wake.  Formally, this traces back to
differences in the effective inertial mass and buoyancy mass of
the particle -- see equation~(\ref{EQR}).  In figure~\ref{tu1ve2},
we show the velocity variation for a light glass sphere (density
2.48) compared to a tungsten bead (density 14.8).  We observe that
the velocity of the glass oscillates before reaching a constant
terminal value whereas the other particle has a regular
acceleration.  In the case of light beads the hydrodynamic forces
may be large enough to overcome the gravity and change the sign of
the acceleration . This is linked with the non-stationarity of the
wake, as vortex shedding is known to occur for Reynolds number
above critical ($Re_{c} \sim 250$).

\subsection{Turbulent flow : Lagrangian velocity measurement}
%-------------------------------------------------------------

\subsubsection{Experimental set-up}
\begin{figure}
    \includegraphics[height=6cm]{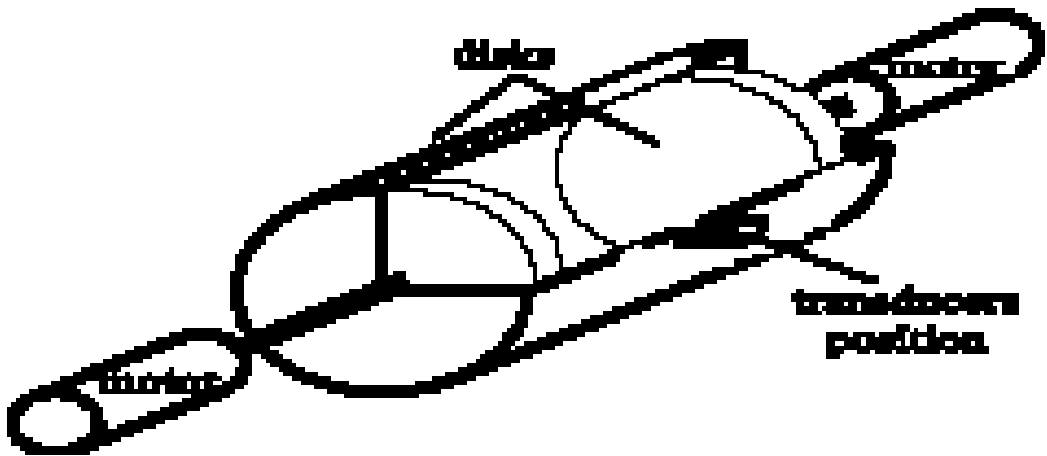}
    \caption{Experimental setup.  The inner radius of the cylinder is 10~cm (disks radius $R$=9.5~cm) and
    the distance between the disks is 18~cm. The disks are
    driven by two 1~kW motors at a constant rotation frequency of
    $f$=18~Hz. The transducers are placed 18~cm off axis, in order to increase the volume of the measurement
    region.}
    \label{klac}
\end{figure}

%----------------------------------
The turbulent flow is generated in a von K\'arm\'an geometry~: the
water is set into motion by two coaxial counter rotating disk in a
cylindrical tank (figure~\ref{klac}).  The Reynolds number
$Re=\frac{2 \pi R^2 f}{\nu}$ (where $\nu=0.89
10^{-6}$~m$^2$s$^{-1}$ is the water kinematic viscosity) is equal
to $10^6$.  To prevent cavitation in the flow, we boil the water
before filling the tank by lowering the pressure with a vacuum
pump and during the experiment the pressure is increased to two
bars. For the acoustic measurement, we use the same array of
transducers as in the previous experiments, at emitting frequency
3~MHz.  The cylinder and the surface of the disks are covered by
3~cm of Ciba Ureol 5073A and 6414B. Its density is 1.1 and the
sound velocity is 1460~m.s$^{-1}$ so that its acoustic impedance
is close to that of the water, reducing drastically the
reflections at the interface water/ureol compared to water/steel.
The attenuation at 2.5~MHz is about 6~dB per cm.  With a 3~cm
layer and after the reflection on steel the total absorption is
about 36~dB. The total reflection at the interfaces is reduced by
a factor 60.  The particle is a polypropylene (PP) sphere of
radius 1~mm and density $0.9$.

\subsubsection{Results}
\begin{figure}
    \centering
    \subfigure
    {
    \rotatebox{90}{\includegraphics[width=6cm]{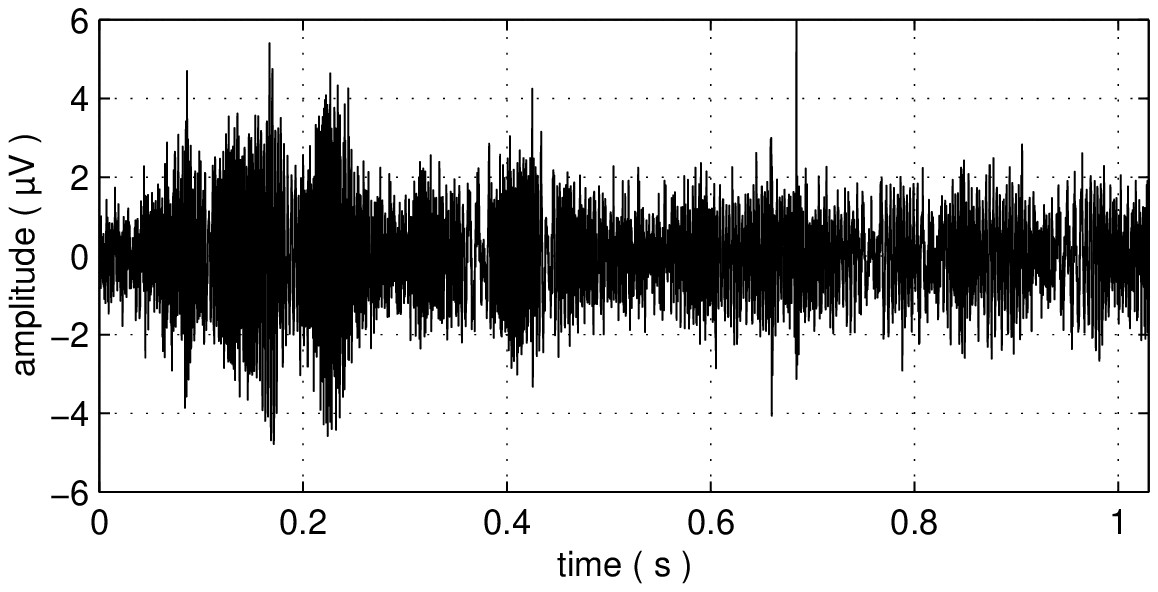}}
    }
    \subfigure
    {
    \rotatebox{90}{\includegraphics[width=6cm]{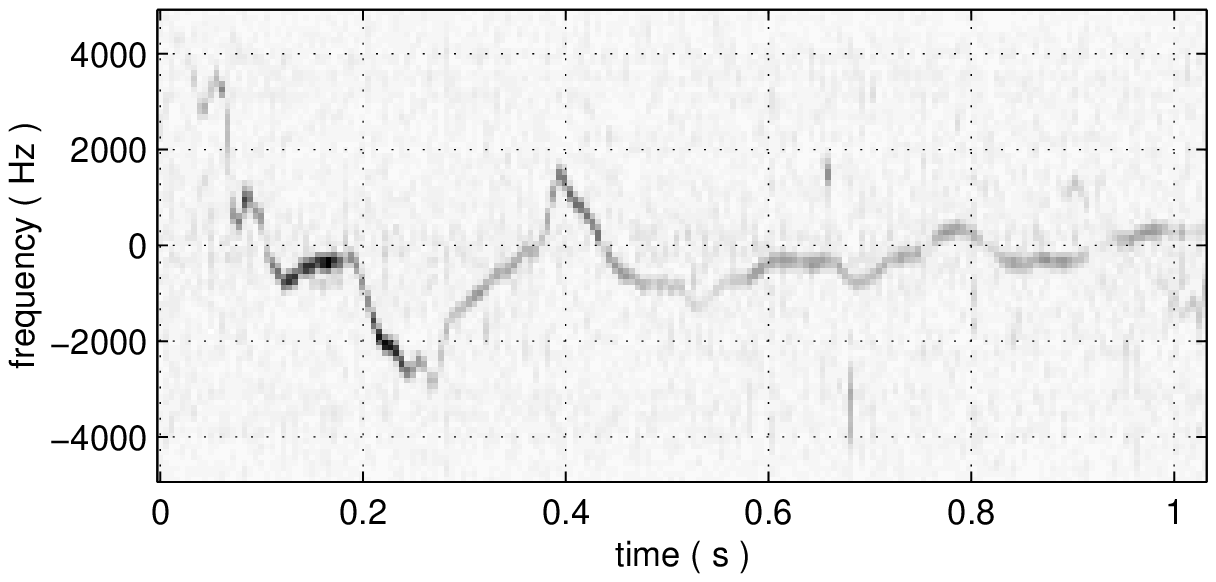}}
    }
    \subfigure
    {
    \rotatebox{90}{\includegraphics[width=6cm]{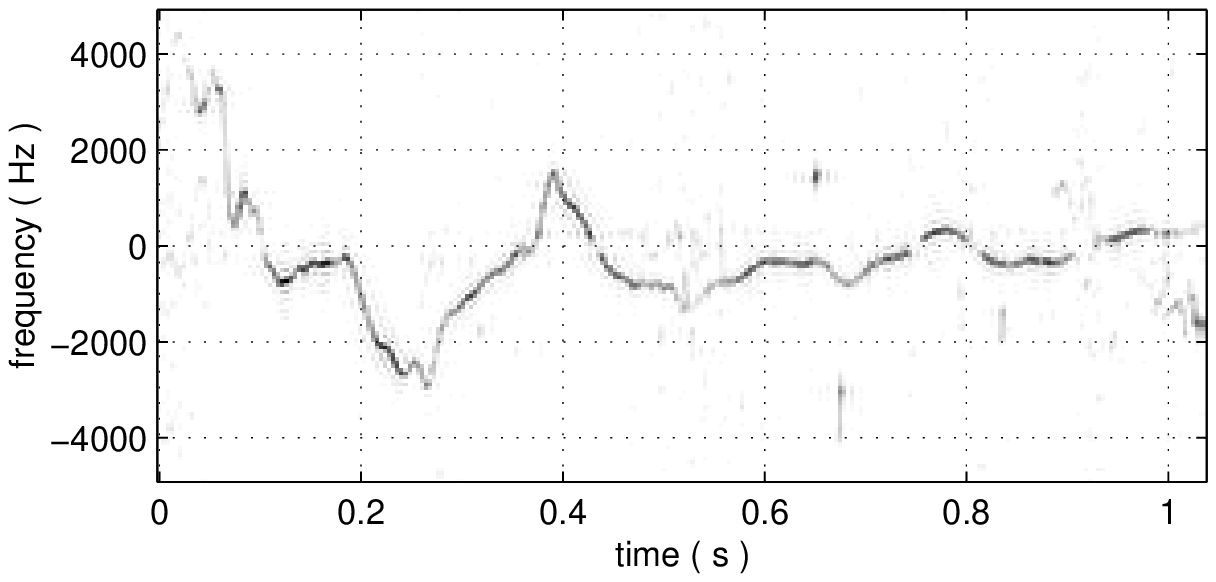}}
    }
    \caption{Sound scattered by a 2~mm diameter PP bead in a turbulent flow at
    $Re=10^6$.
    (a) Typical time series; (b) and (c) corresponding
     spectrogram and reassigned spectrogram.}
    \label{sigklac}
\end{figure}

%-----------------------
We show in figure~\ref{sigklac} the time series when one particle
is in the ultrasonic beam and the corresponding spectrogram and
reassigned spectrogram. The signal to noise ratio is very poor,
typically less than 6dB (to give an idea, in
figure~\ref{sigklac}(a) the bead enters the ultrasonic beam at $t
\sim 20$~ms). One can also see some events localized in the time
frequency plane that may be considered as noise and that may have
several origin (noise of the motors, external electromagnetic
noise, cavitation in the flow\ldots). Altogether, the
time-frequency pictures show the trajectory of the particle but
the low SNR prevents it from being easily extracted. In
particular, the trajectory in the reassigned picture becomes quite
lacunar and extracting it would require sophisticated (and CPU
greedy) image processing techniques.

\begin{figure}
   \centering
    \subfigure
    {
    \rotatebox{90}{\includegraphics[width=6cm]{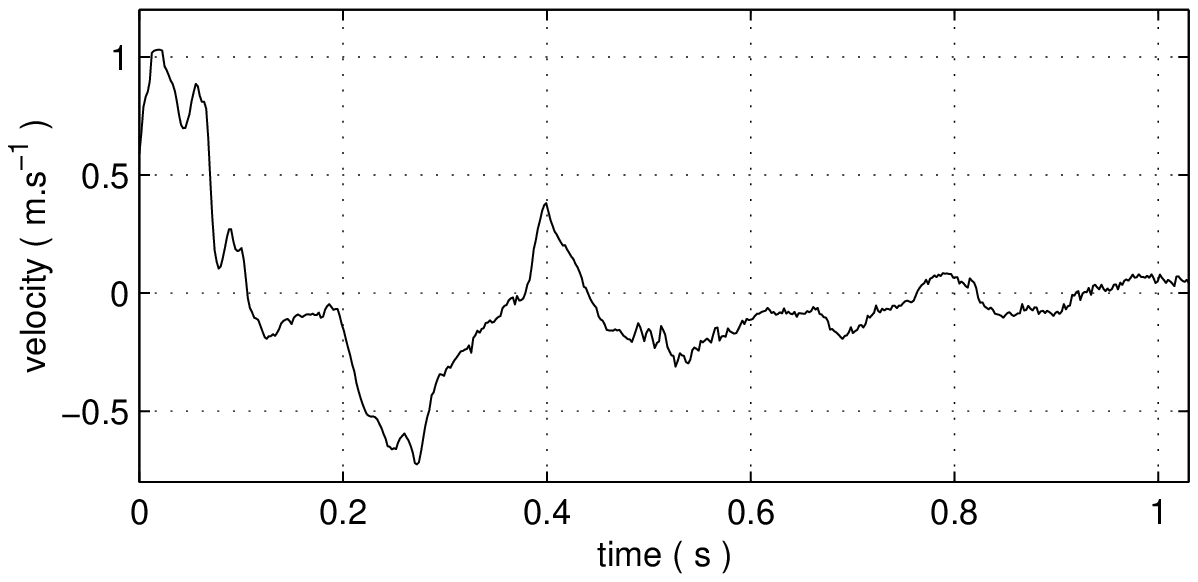}}
    }
    \subfigure
    {
    \rotatebox{90}{\includegraphics[width=6cm]{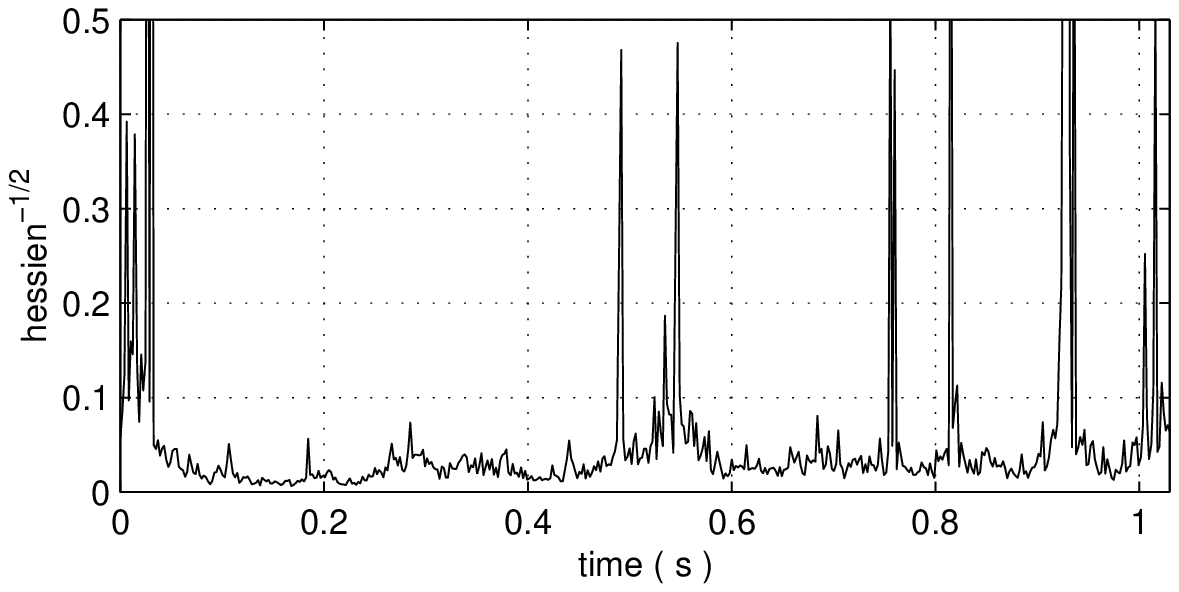}}
    }
    \subfigure
   {
    \rotatebox{90}{\includegraphics[width=6cm]{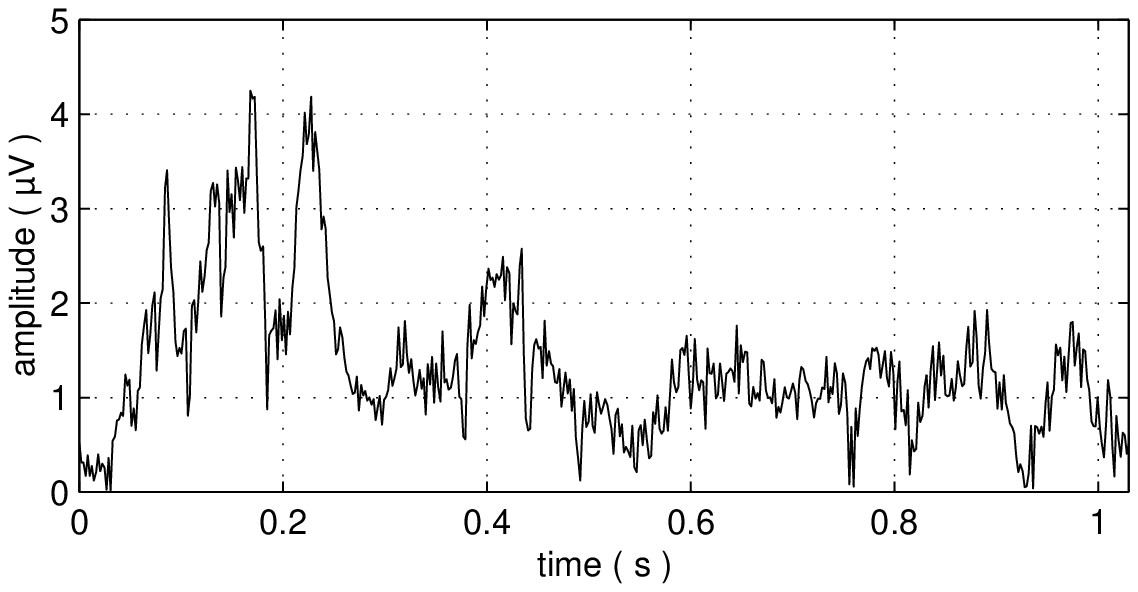}}
   }
   \caption{Velocity measurement for the motion of a 2~mm diameter PP bead in a turbulent flow at
    $Re=10^6$. Output of the AML algorithm: (a) velocity
   (b) corresponding inverse square root of the hessian, (c) amplitude of the source (the rms value of the noise is 0.9$\mu$V).
   AML algorithm parameters: M=1, K=7, Q=13, v$^2$=10$^{-5}$.}
   \label{vithes}
\end{figure}

The result of the AML algorithm is plotted in figure~\ref{vithes}.
The extracted frequency modulation is of course within the
estimation in the spectrogram as in figure~\ref{sigklac}(b), but
one observes that fine variations in the velocity of the bead are
now detected. The algorithm provides also an estimate of the
amplitude of the source (figure~\ref{vithes}(c)). It can be seen
that there is a strong amplitude modulation and that the SNR is at
most 6~dB and may become less than 0~dB.

As the hessian is related to the Fisher information
matrix~\cite{Scharf}, its inverse square root is linked with the
variance of the estimation: a large value of the hessian indicates
an accurate estimation of the modulation frequency and, hence, of
the bead velocity. The inverse square root of the hessian is
plotted in figure~\ref{vithes}(b): very large values are
calculated in the absence of a bead in the measurement volume at
the beginning and end of the time series (as a signature of the
mismatch between the model which is composed of one source at
least and the reality: no source). Local lower values (typically
less than 0.1) are observed when the variance on the estimation is
small. Spurious effects are generated when the frequency
modulation approaches zero as the hessian also becomes very small
because of the filtering operation made in order to get rid of the
coupling part of the signal. Finally, one observes that the
hessian decreases as the signal to noise ratio increases (see at
time 0.55~s).

\section{Concluding remarks}
%---------------------------

As can be seen in the previous section, both methods,
time-frequency analysis and parametric spectral analysis are
suited for extracting the time-varying frequency modulation due to
a Doppler effect. The domain of application of each method depends
on the degree of non-stationarity and on the SNR.

For high SNR and weakly non-stationary signals, the time-frequency
approach yields very good results.  One drawback is the need of a
second processing stage to extract the trajectory from the
time-frequency picture.  This stage may become increasingly
difficult if there is more than one spectral component or if the
SNR degrades. In both cases the quadratic nature of the algorithm
produces interference patterns in the image: spurious clusters and
a lacunar trajectory result. Another, more fundamental, limitation
is that the length of the time window must be long enough to
preserve an acceptable frequency resolution, even with the
reassigned spectrogram. This limits the methods to weakly
non-stationary signals.

For signals with a rapid frequency modulation, the AML spectral
estimation is well suited, as long as the noise is near iid. The
size of the time window can be decreased because of the parametric
nature of the method, since {\it a priori} knowledge has been
taken into account. The performance is further increased by the
use of a Kalman-like filter. The drawback is the necessity to find
a good dynamical model for the evolution of the spectral
components.  We have chosen here the simplest model which works
well for our experiments but the approach can be refined by
increasing the number of parameters in order to consider more
precisely the variation of the frequency. The AML algorithm also
provides a quantitative estimation of the quality of the
demodulation and the instantaneous power of the spectral
component.  Finally, the AML method has the advantage to provide
directly frequency modulation as a function of time, in one stage.

\begin{acknowledgments}
We are indebted to Pascal Metz for the development of the signal
conditioning electronics. We thank Marc Moulin for his help in the
design of the von K\'arm\'an setup, VERMON for continuous
assistance in the development of the transducer array. This work
is partially supported by ACI grant No. 2226.
\end{acknowledgments}

\appendix

\section*{AML algorithm}

\begin{itemize}

\item
First step : calculate $\hat \MR_x$ using the following
expression\cite{Michel91b} :
\begin{equation}
    \hat \MR_x=\frac{1}{2Q}\sum_{i=t+1}^{t+Q}\left( \X(i)\X(i)^T +
    \tilde{\X}(i)\tilde{\X}(i)^T \right) \; ,
\end{equation}
with
\begin{equation}
    \tilde{\X}(i)={[x(i+K-1),x(i+K-2),\dots,x(i)]^{*}}^T \; ,
\end{equation}
where $^*$ stands for complex conjugate. $\tilde{\X}$ is the
complex conjugate of the time reversed version of $\X$.

\item
Second step : diagonalize $\hat \MR_x$; one obtains the
eigenvectors $(\V_i)_{i=1..K}$ and eigenvalues
$(\lambda_i)_{i=1..K}$ sorted in decreasing order.

\item
Third step : Compute $\hat \Pi_n$ and $\hat \sigma ^2$, using the
set of equations
\begin{eqnarray}
   \hat\Pi_n=\sum_{i=M+1}^{K}\V_i\V_i^T \; , \\
   \hat \sigma^2 = \frac{1}{K-M}\Tr(\hat \Pi_n \hat
   \MR_x)=\frac{1}{K-M} \sum_{i=M+1}^K \lambda_{i} \; .
\end{eqnarray}

\item
Forth step : choose $\F=\hat{\F}(t)$ as candidate value. Compute
${\vec{\rm grad}}$ and $\MH$ using\cite{Michel91a}
\begin{eqnarray}
{\vec{\rm grad}}= \frac{2Q}{\sigma^2} {\rm R_e} \left\{ {\rm
Diag}(\MS'^+(\F).\Pi_n(\F).\hat\Pi_n.\MS(\F).\hat\MP) \right\} \;
,\\ \MH=\frac{2Q}{\sigma^2}{\rm R_e}\left\{ {\rm Diag} \left(
(\MS'^+(\F).\Pi_n(\F).\hat\Pi_n.\Pi_n(\F).\MS') \right)
\star\hat\MP^*\right\} \; ,
\end{eqnarray}
where the operator $\star$ stand for the term to term matrix
multiplication, and $\MP^*$ is the conjugate of $\MP$, and \beq
   \MS' = [\frac{d\vS_1}{df_1}, \ldots,\frac{d\vS_M}{df_M}]^T \; .
\eeq

\item
Fifth step : using equations (\ref{eqa}) to (\ref{eqd}), compute
$\hat{\F}(t+1)$ and $\Gamma(t+1)$.
\end{itemize}

The initialization of the algorithm is done by either (i) setting
an initial value of $\F(1)$ or (ii) estimating this value using
the maxima of the amplitude of the FFT of a small window of signal
(of length 64 or 128 samples) and using the iterative algorithm
described in section \ref{AMLsec} to converge towards $\F(1)$.

For example, the extracted velocity of figure~\ref{vithes} is
obtained by starting at the maximum of energy of the signal and
applying the algorithm forward and backward in time. The algorithm
is stopped as the mean of the inverse square root of the hessian
over a window of size 400 samples exceeds 0.5 for more than 400
samples.

\bibliography{jasa}
\end{document}